\documentclass[11pt]{article}

\usepackage[a4paper,margin=3cm]{geometry}

\usepackage{authblk}

\usepackage[parfill]{parskip}
\usepackage{setspace}

\usepackage[labelfont=bf,font={small,stretch=0.95}]{caption}

\usepackage[square,super,sort,compress,numbers]{natbib}
\usepackage{hyperref}

\usepackage{gensymb}
\usepackage{amsmath}
\usepackage{amssymb}

\usepackage{graphicx}

\usepackage{multirow}

\usepackage[version=4]{mhchem}

\begin{document}

\title{\LARGE\bf Accelerated First-Principles Exploration of Structure and Reactivity in Graphene Oxide}

\author[1]{Zakariya El-Machachi}
\author[1]{Damyan Frantzov}
\author[1]{A. Nijamudheen}
\author[2]{Tigany Zarrouk}
\author[2]{Miguel A. Caro}
\author[1]{Volker L. Deringer\thanks{volker.deringer@chem.ox.ac.uk}}

\affil[1]{Inorganic Chemistry Laboratory, Department of Chemistry, University of Oxford, Oxford OX1 3QR, United Kingdom}
\affil[2]{Department of Chemistry and Materials Science, Aalto University, 02150 Espoo, Finland}

\date{}

\maketitle

\begin{abstract}
Graphene oxide (GO) materials are widely studied, and yet their atomic-scale structures remain to be fully understood. Here we show that the chemical and configurational space of GO can be rapidly explored by advanced machine-learning methods, combining on-the-fly acceleration for first-principles molecular dynamics with message-passing neural-network potentials. The first step allows for the rapid sampling of chemical structures with very little prior knowledge required; the second step affords state-of-the-art accuracy and predictive power. We apply the method to the thermal reduction of GO, which we describe in a realistic (ten-nanometre scale) structural model. Our simulations are consistent with recent experimental findings and help to rationalise them in atomistic and mechanistic detail. More generally, our work provides a platform for routine, accurate, and predictive simulations of diverse carbonaceous materials.
\end{abstract}

\setstretch{1.5}

Graphene oxide (GO) is a summary term for a range of layered materials created by reacting graphite with aggressive agents, such as \ce{KMnO4}, typically followed by partial reduction and sometimes functionalisation.\cite{dimiev_graphene_2017, dreyer_chemistry_2010, guo_controlling_2022, wu_graphene_2023} Today, GO materials can be controllably prepared\cite{dikin_preparation_2007} and find emerging applications in catalysis,\cite{su_probing_2012} membranes,\cite{joshi_precise_2014} electronics,\cite{eda_chemically_2010} and photonics.\cite{wu_graphene_2021} Despite decades of work, however, the precise chemical structure of these materials has remained elusive. The ordered regions of GO sheets can be directly visualised using high-resolution electron microscopy,\cite{erickson_determination_2010, dave_chemistry_2016} but the nature of the more disordered regions can only be inferred from indirect observations, such as vibrational and nuclear magnetic resonance (NMR) spectroscopy. The properties of GO materials cannot be unambiguously linked to chemical structure if this structure itself is not precisely known (which functional groups are present; in what amounts?).

To complement experimental techniques, GO has been widely studied by computational chemistry methods. For example, Kumar et al.\ combined reactive-force-field simulations with density-functional theory (DFT) to show how varying functional groups affect the stability and electronic structure of thermally reduced graphene oxide (rGO), and how rGO forms graphitic and oxidised domains during thermal annealing.\cite{kumar_impact_2013,kumar_scalable_2014} Atomistic modelling of rGO further revealed that defects formed during thermal reduction can lead to pores for applications in water desalination and natural gas purification.\cite{lin_atomistic_2015} Explicit water molecules have been incorporated into computational models of GO membranes to simulate interlayer separation and water diffusivity, providing insights for applications.\cite{williams_computational_2018, williams_silico_2019,Futamura2024} Theoretical studies delved into aspects such as the excess surface charge in hydrated GO and the dynamic evolution of functional groups, employing \textit{ab initio} molecular dynamics (AIMD) for a comprehensive understanding of GO in water.\cite{mouhat_structure_2020} 

Despite these advances, there remains an inherent limit to the length and time scales accessible to AIMD. Machine learning (ML) based interatomic potentials provide an emerging alternative approach that promises much faster simulations while retaining quantum-mechanical accuracy.\cite{Behler2017, deringer_mlip_2019, Friederich_ff_2021, unke_ff_2021} In the context of carbon materials, ML-driven simulations have been used to describe defective\cite{thiemann_defect_2021} and fully amorphous graphene,\cite{el-machachi_exploring_2022} the growth of carbon thin films,\cite{caro_growth_2018} and the formation of voids in low-density porous forms.\cite{deringer_towards_2018, wang_structure_2022, ugwumadu_self_2023}

In the present work, we show how one can rapidly explore a wide range of functional groups and disorder in GO materials by combining two recent innovations in atomistic ML (Figure 1). First, we use on-the-fly-accelerated AIMD \cite{Li2015,Jinnouchi2020,stenczel_castep_gap_2023} to efficiently sample configurations for an initial training dataset. Second, we show that this approach can be used to kick-start a much wider exploration using state-of-the-art neural-network potentials. For the first task, we use CASTEP+ML\cite{clark_castep_2005, stenczel_castep_gap_2023} coupled to the Gaussian approximation potential (GAP) ML framework;\cite{bartok_gap_2010, bartok_soap_2013, deringer_gap_review_2021} for the latter, we use an equivariant neural-net\-work architecture based on the mes\-sage-passing atomic cluster expansion (MACE).\cite{batatia_mace_2022, batatia_design_2022, kovacs_evaluation_2023} A key point of our study is that those two principally different methodologies can be synergistically combined. The predictions of the final ML model agree remarkably well with experimental observations, showing promise for future applications to the chemistry of carbon-based materials.

\begin{figure*}[t]
    \centering
    \includegraphics[width=\textwidth]{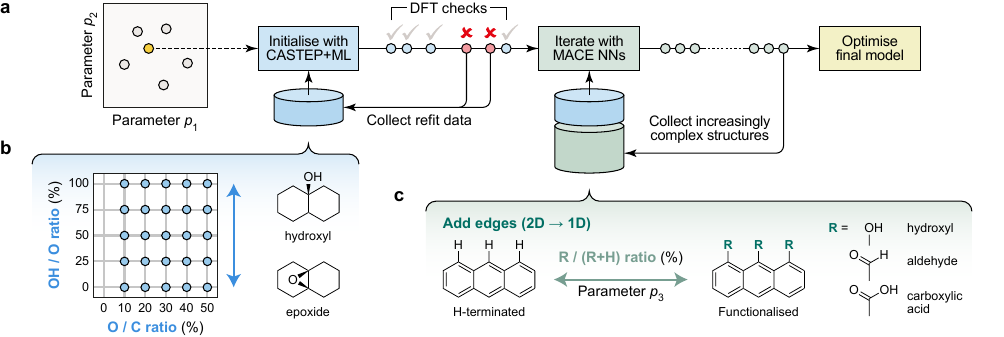}
    \caption{Accelerated exploration of functional groups in GO with machine-learning-driven simulations. (a) Schematic overview of the approach. We initialise the search for structures with CASTEP+ML trajectories, which combine first-principles MD with on-the-fly fitting of ML potentials \cite{stenczel_castep_gap_2023} (blue). Once this initial generation is complete, iterative training kicks in, exploring increasingly complex structural spaces (green). The data are then used to train and optimise the final model (yellow). (b) Parameter space of 2D functionalised GO, with a schematic sketch of how the OH / O ratio controls the ratio of hydroxyl and epoxy groups in the initial structures. (c) Extension of the parameter space to include 1D structures (edges and ribbons), which can be hydrogen-terminated or functionalised with different groups, R. }
    \label{fig:parameter_space}
\end{figure*}

To sample the wide variety of possible GO structures, we define a space of \textit{N} parameters, $\mathcal{P} = \left[p_{1}, ..., p_{N}\right]$, that determines the composition of an initial candidate structure. Here, in line with existing knowledge in the field, \cite{dimiev_graphene_2017, dreyer_chemistry_2010, guo_controlling_2022, wu_graphene_2023} we choose these parameters to be: (1) the ratio of \ce{O} to \ce{C} atoms in the initial sheet, determining the degree of oxidation; (2) the \ce{OH} / \ce{O} ratio, {\em i.e.}, the concentration of hydroxyl groups relative to all \ce{O} atoms; (3) the ratio of functionalised edges (--OH, --CHO, or --CO$_{2}$H) to hydrogen-terminated ones. We explored up to $p_{2}$ in CASTEP+ML runs (that is, we functionalised only 2D graphene sheets), and up to $p_{2}$ and then $p_{3}$ in MACE iterations. 

We started the process with 25 CASTEP+ML runs, at 300~K for 10 ps each, corresponding to the grid shown in Figure 1b. Most of those simulations (20 of 25) ran to completion; five terminated early due to erroneously lost atoms. The latter results are still valuable, as they contain high-energy and -force structures which can be used to guide early models away from unphysical configurations during iterations.  
In all, 820 CASTEP+ML structures were used for the initial training dataset.
Next, equivariant MACE potentials were trained in an iterative fashion that gradually extended the scope of the model: 
(i) exploring higher temperatures in MD runs, gradually increasing from 600 to 1,500 K; 
(ii) repeating the protocol at 1,500 K for the next four iterations;
(iii) finally, exploring 1D structures at 1,500 K.
After a total of 12 iterations, training structures with any force component $>50$ eV/Å  were filtered out to further enhance the model. The final training dataset contains 3,016 simulation snapshots (605,204 atoms).

\begin{figure}[t]
    \centering
    \includegraphics[width=8.5cm]{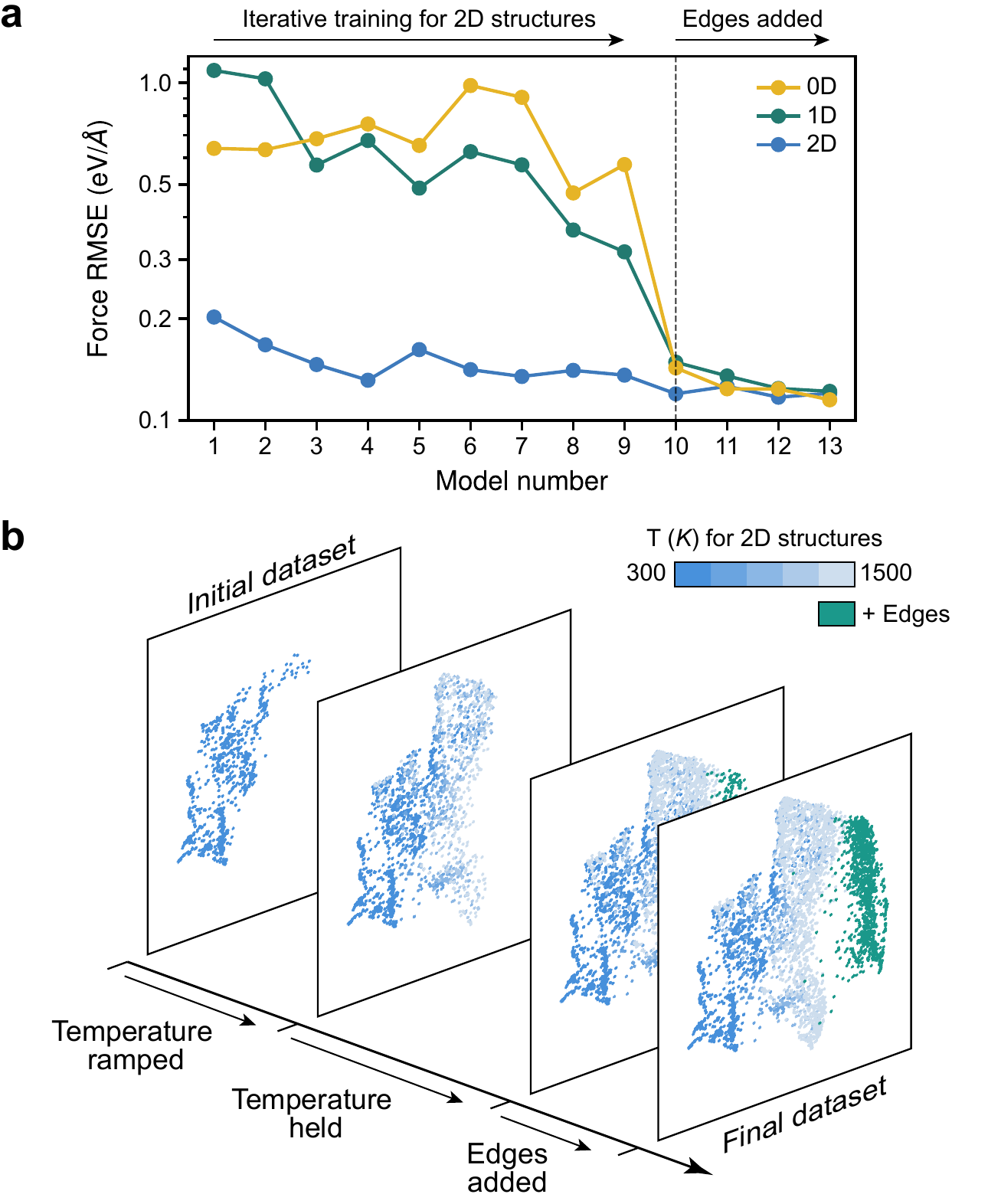}
    \caption{Training ML models for GO. (a) Root-mean-square error (RMSE) of forces predicted by iteratively trained MACE models. Errors are evaluated on an external set of DFT data not included in the training, comprising 100 snapshots each of a 2D sheet (blue) and a 1D nanoribbon (green) sampled in separate AIMD trajectories at 300 K, as well as 0D nanoflake structures (yellow) taken from Ref.~\citenum{motevalli_representative_2019} and re-evaluated at the relevant level of DFT. (b) Structural diversity during iterative training. The different iterations are visualised by UMAP embedding\cite{McInnes2018} of a kernel-based structural similarity metric.\cite{bartok_soap_2013} Points are colour-coded according to the temperature set in the MD simulation for 2D structures (blue), whereas they are shown in a single different colour for 1D edge structures (green).}
    \label{fig:learning-curves-and-maps}
\end{figure}

We test the performance of our ML models on external data not seen in training: two AIMD trajectories of functionalised 2D sheets and 1D nanoribbons, respectively, as well as single-point calculations for molecular (0D) fragments taken from Ref.~\citenum{motevalli_representative_2019}. 
Figure 2a shows how the force error -- our main performance metric -- evolves during iterations. As more data are added, the errors decrease for the 2D test set, as expected. 
For 1D and 0D structures, the errors are initially high since early models have not ``seen'' edges, specifically C--H bonds which are explicitly included only from model 10 onwards. Adding edge structures rapidly reduces the corresponding errors (dashed line in Figure 2a). The final force accuracy is similar across all benchmarks, just over 0.1 eV/\AA{}.

To illustrate the gradual exploration of chemical and configurational space, we show four SOAP similarity maps \cite{bartok_soap_2013, Cheng2020} in Figure 2b: the CASTEP+ML seed at 300 K, the dataset after gradually ramping to 1,500 K, the inclusion of the first edge structures, and the final dataset. The initial structures form two distinct clusters on the map; at higher $T$, one cluster grows and a third, smaller one appears. Finally, including edges adds a distinct set of structures (green).

We now describe an application of the final MACE model to a challenging problem in materials chemistry -- name\-ly, to large-scale MD simulations of the thermal reduction of GO to rGO. This process involves a vast number of functional groups which transform and eventually disappear, accompanied by the evolution of gaseous species such as \ce{CO2}. Experimentally, reduction temperatures of 1,100 $\degree$C yielded resistivity values of $\sim10^{-5} \text{ }\Omega$ m,\cite{mattevi_evolution_2009} on par with that of graphite.\cite{tyler_thermal_1953} Understanding how functional groups evolve during thermal reduction could help to correlate the structure of the sheet with its properties. We show in the following that our ML-accelerated approach can provide such an atomic-scale understanding.

\begin{figure*}[t]
    \centering
    \includegraphics[width=\textwidth]{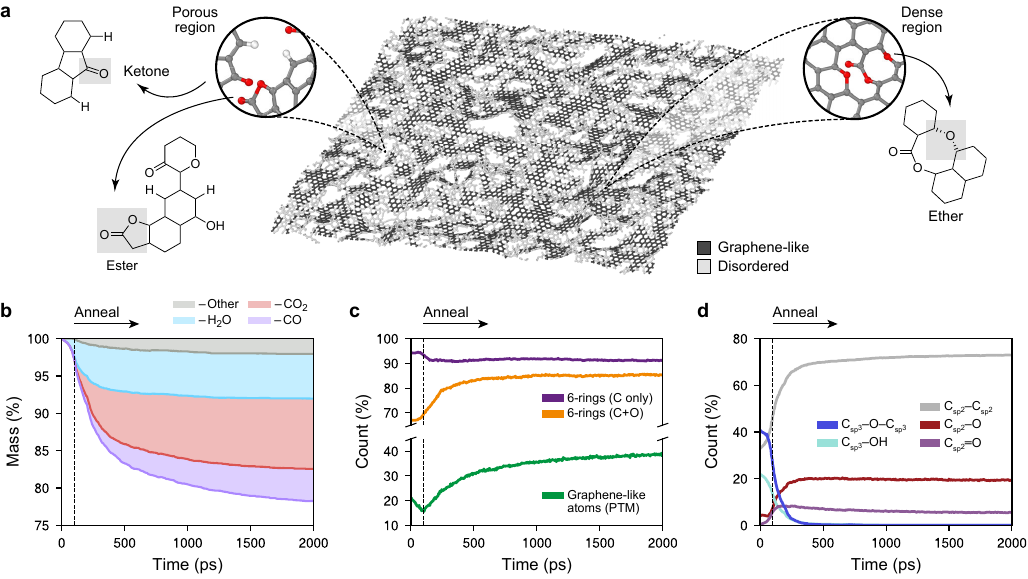}
    \caption{A large-scale structural model of reduced graphene oxide (rGO) generated through simulated thermal reduction. (a) The rGO structure after 1.5 ns of simulation time (3 million timesteps). The dark grey regions highlight graphene-like regions identified using polyhedral template matching (PTM).\cite{larsen_ptm_2016, OVITO} Insets show close-ups with C atoms in grey, O in red, and H in white. (b) The change in mass of the GO sheet as it is thermally reduced in the 1,500 K simulation. The three most common leaving molecules (\ce{H2O}, blue; \ce{CO2}, red; \ce{CO}, purple) are tracked in the stacked plot, along with other species (grey). (c) Evolution of structural indicators: the count of 6-membered rings (shown separately for only C-based rings, purple, and for all rings, orange), and the count of graphene-like atoms identified by PTM. (d) Evolution of functional groups bonded to sp${^2}$ and sp${^3}$ carbon atoms, respectively, obtained using a topological bond-counting algorithm. }
    \label{fig:rGO}
\end{figure*}

Our starting structure is a partially disordered, fully sp$^{2}$-bonded graphene sheet with 10,368 atoms ($17.7 \times 15.3$ nm$^{2}$ in a single layer). The sheet was generated using Monte-Carlo bond switching driven by ML local-environ\-ment energies, following Ref.~\citenum{el-machachi_exploring_2022}, and then functionalised with $\mathcal{P} = [0.4, 0.5, 0]$, raising the atom count to 16,645. This structural model represents features of GO including the topological disordering of the carbon backbone (presence of non-6-membered rings), although it is constrained to three-fold coordination for all carbon atoms, and therefore does not initially contain large pores. 

The thermal reduction was studied in three independent MD simulations at temperatures of 900, 1,200, and 1,500 K, respectively. The structures were rapidly heated over 100 ps and then held at the respective annealing temperature for 1.9 ns. We note that experimental protocols for thermal reduction of GO span a wide range of parameters: temperatures from 80 to 1,100 $\degree$C\cite{mattevi_evolution_2009} and times from 10 minutes\cite{maphiri_novel_2021} to 5 days.\cite{kumar_scalable_2014} Computationally, we are limited by the timescales accessible to MD (on the order of nanoseconds); thus, more aggressive heating is used to overcome local energy barriers. \cite{DeTomas2017}
We found that annealing at 1,500 K yields a structure in good agreement with experiment, which we discuss below; results for the other MD runs are given in the Supporting Information.

Figure 3a shows the rGO structural model during annealing at 1,500 K. ``Graphene-like'' regions, shown in dark grey, form small islands embedded within disordered and porous regions. (We quantify ``graphene-like'' content through polyhedral template matching, a method to identify crystal-like local environments.\cite{larsen_ptm_2016})  This result agrees qualitatively with electron microscopy images clearly showing amorphous regions together with holes and pores in the structure.\cite{Reyes2022} 

The formation of this structure is accompanied by a mass loss of $>20$\% as gaseous species leave the surface (Figure 3b), which can be qualitatively correlated with thermogravimetric experiments.\cite{valentini_tuning_2023} Initially, as the temperature ramps up over the first 100 ps, nearly all mass loss is due to \ce{H2O} (light blue shading in the stacked plot of Figure 3b).  Having reached 1,500 K (dashed line), the first \ce{CO2} molecules detach -- and this species, indicated by red shading, quickly begins to dominate the mass loss as the sheet is reduced. \ce{CO} was also evolved in notable amounts (magenta); other gaseous species such as \ce{OH}, \ce{C3O2}, \ce{H2O2}, etc., were occasionally observed but were mostly rare and short-lived.   

\begin{figure}[t]
    \centering
    \includegraphics[width=8.5cm]{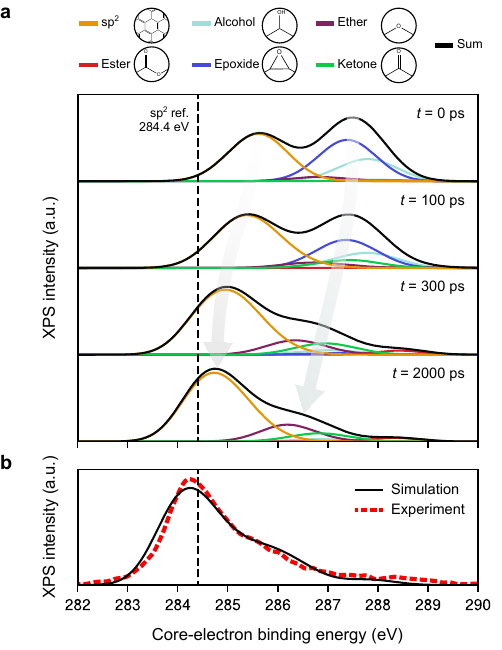}
    \caption{X-ray photoelectron spectroscopy (XPS) predictions for the rGO structure annealed at 1,500 K. (a) ML-predicted XPS spectra at different points of the simulation. The grey arrows indicate a clear shift to lower core-electron binding energy values and also a reduction in the second peak during the simulation as oxygen-based functional groups are removed. The vertical, dashed black line indicates the reference sp$^{2}$ value at 284.4 eV. (b) XPS prediction for the final structural model obtained after full structural optimisation. The simulation data in panel (b) are shifted horizontally to align with the experimental data from Ref.~\citenum{valentini_tuning_2023}.}
    \label{fig:GO-anneal}
\end{figure}

In addition to the mass loss, our simulations allow us to address the changes in the structure of the GO sheet itself. Figure 3c shows the fraction of ``graphene-like'' atoms in ordered local environments: initially, their percentage decreases, correlating with the loss of \ce{H2O}; then, as \ce{CO2} is released, there is a clear and concomitant increase in graphenic content.  These observations agree with experimental findings where \ce{CO2} and, to a lesser extent, \ce{CO} loss leads to defects in the GO sheet,\cite{valentini_tuning_2023, Nguyen_rgo_2015, Pelaez_2021} inducing structural rearrangements and increasingly more graphene-like environments. Interestingly, the count of 6-membered rings containing only carbon decreases slightly (purple in Figure 3c), whereas if we consider all atoms in the sheet in the ring analysis, the 6-membered rings increase from just above 65\% to $\approx 85$\% (orange). This analysis suggests that oxygen atoms take part in substantial rearrangements during annealing, from sp$^{3}$ environments perpendicular to the basal plane such as epoxide and alcohol groups, to sp$^{2}$ environments parallel to the plane -- for example, ethers and esters (cf.\ Figure 3a). 

Beyond the overall ``graphene-like-ness'', we can also trace the evolution of individual functional groups, enabling qualitative comparison with $^{13}$C NMR results from Ref.~\citenum{valentini_tuning_2023}. Figure 3d shows how the C$_{\text{sp}^{2}}$–C$_{\text{sp}^{2}}$ count increases exponentially during the initial 100 ps -- in contrast to the graphenic content, which decreases during this period (Figure 3c). This observation suggests that as water molecules leave the sheet, the carbon backbone will initially form defective rather than ordered sp$^{2}$ environments. 
Then, during high-$T$ annealing, the sp$^{2}$ count grows more gradually, indicating a transformation to a more graphene-like carbon backbone, consistent with Figure 3c. The loss of oxygen-based molecules (cf.\ Figure 3b) is clearly mirrored in a declining C$_{\text{sp}^{3}}$–O–C$_{\text{sp}^{3}}$ (epoxide) and C$_{\text{sp}^{3}}$–OH (hydroxyl) count during the first 300 ps of the simulation. Concomitantly, C$_{\text{sp}^{2}}$–O groups form,
which has been proposed as a mechanism by which the material drops in resistivity without a significant change in mass.\cite{valentini_tuning_2023} Once the sp$^{3}$-bonded epoxide and hydroxyl groups are removed, the C$_{\text{sp}^{2}}$–O count remains at an almost steady state. Finally, the C$_{\text{sp}^{2}}$=O (carbonyl) count peaks at $\approx 250$ ps before gently decreasing for the remainder of the annealing simulation. 

With a quantum-mechanically accurate description of the chemical structure in hand, the newly created structural models can now be analysed with advanced X-ray spectroscopy predictions,\cite{golze_accurate_2022, Zarrouk2024experimentdriven} as we have previously exemplified for small-scale GO models (with DFT-level predictions at the time).\cite{aarva_xray_2021} Here, applying the ML spectroscopy model of Ref.~\citenum{golze_accurate_2022} reveals two peaks in the predicted X-ray photoelectron spectroscopy (XPS) data for the initial structure (Figure 4a). The first peak corresponds to unmodified ``sp$^{2}$'' carbon atoms, whilst the second relates to oxygen/hydrogen-based functional groups, with the largest contribution arising from  C$_{\text{sp}^{3}}$–O–C$_{\text{sp}^{3}}$ (epoxide, dark blue) and C$_{\text{sp}^{3}}$–OH (alcohol, light blue) groups. The core electron binding energies (CEBEs) of all functional groups are shifted upwards from experimental reference energies due to the electronegativity of oxygen. The eventual removal of these oxygen-based groups during annealing reduces the magnitude of the aforementioned CEBE shifts: all motifs decrease in CEBE.

Comparison with experimental data from Ref.~\citenum{valentini_tuning_2023}, shown in Figure 4b, reveals good agreement between theory and experiment for rGO at 300 K (in air). Experimentally, XPS spectra use a fixed reference value for deconvolution, which does not take into account the effect of local interactions from electronegative species. As a result, the spectra are shifted accordingly to align well with experimental data. We refer the reader to Refs.~\citenum{golze_accurate_2022} and \citenum{Zarrouk2024experimentdriven} for a detailed discussion on this matter. 

In conclusion, we have reported a computational approach to modelling and understanding the highly diverse chemical structure of GO, and we have shown an initial application to the thermal reduction of this material. Our work combines two recent developments in atomistic ML.
For CASTEP+ML, we view its main advantage in this context to be in saving ``human time'': it allows the researcher to create, from scratch, a chemically diverse training dataset to seed a new ML potential with minimal manual input.\cite{stenczel_castep_gap_2023}
For MACE, our work builds on recent capability demonstrations, \cite{Kovacs2023a, Batatia2023} showing that this architecture can be combined with efficient dataset-building workflows to readily deploy to new, challenging modelling problems in chemistry.

Looking forward, we expect this combined methodology to provide a powerful platform for further studies of GO materials. 
For example, the present proof-of-concept for XPS prediction during (simulated) structural transformations could provide a motivation for future {\em in situ} experiments.
Beyond the simulations of GO in vacuum reported here, the interaction of the material with water has been studied using empirical\cite{williams_computational_2018, williams_silico_2019} and DFT methods,\cite{mouhat_structure_2020} and it would now be interesting to use the new ML-accelerated framework to explore the nature of water between GO sheets – building on combined experimental and simulation studies in this area,\cite{zheng_swelling_2017} and also on recent ML-driven work on unconventional phases of water “sandwiched” between sheets of pristine graphene.\cite{kapil_first_2022}

\section*{Acknowledgements}

We thank I.~Batatia for technical discussions about MACE model fitting and code implementation, and Dr C.~Ben Mahmoud for valuable feedback on the manuscript.
We are grateful for support from the EPSRC Centre for Doctoral Training in Theory and Modelling in Chemical Sciences (TMCS), under grant EP/L015722/1. This paper conforms to the RCUK data management requirements. This work was supported by the Engineering and Physical Sciences Research Council [grant number EP/V049178/1]. We are grateful for computational support from the UK national high performance computing service, ARCHER2, for which access was obtained via the UKCP consortium and funded by EPSRC grant ref EP/X035891/1.
T.Z. and M.A.C. acknowledge financial support from the Research Council of Finland under grants no.\ 330488, 347252 and 355301, as well as computational resources from CSC (the Finnish IT Center for Science) and Aalto University's Science IT project.

This research was funded in whole or in part by the Engineering and Physical Sciences Research Council [grant numbers EP/L015722/1, EP/V049178/1, and EP/X035891/1]. For the purpose of open access, the authors have applied a Creative Commons Attribution (CC BY) licence to any Author Accepted Manuscript version arising.

\setstretch{1}


\begin{thebibliography}{10}

\bibitem{dimiev_graphene_2017}
A.~M. Dimiev, S.~Eigler (Editors), \emph{Graphene {Oxide}: {Fundamentals} and {Applications}}, John Wiley \& Sons, Chichester, 1st edition \textbf{2017}.

\bibitem{dreyer_chemistry_2010}
D.~R. Dreyer, S.~Park, C.~W. Bielawski, R.~S. Ruoff, \emph{Chem. Soc. Rev.} \textbf{2010}, \emph{39}, 228.

\bibitem{guo_controlling_2022}
S.~Guo, S.~Garaj, A.~Bianco, C.~Ménard-Moyon, \emph{Nat. Rev. Phys.} \textbf{2022}, \emph{4}, 247.

\bibitem{wu_graphene_2023}
J.~Wu, H.~Lin, D.~J. Moss, K.~P. Loh, B.~Jia, \emph{Nat. Rev. Chem.} \textbf{2023}, \emph{7}, 162.

\bibitem{dikin_preparation_2007}
D.~A. Dikin, S.~Stankovich, E.~J. Zimney, R.~D. Piner, G.~H.~B. Dommett, G.~Evmenenko, S.~T. Nguyen, R.~S. Ruoff, \emph{Nature} \textbf{2007}, \emph{448}, 457.

\bibitem{su_probing_2012}
C.~Su, M.~Acik, K.~Takai, J.~Lu, S.-j. Hao, Y.~Zheng, P.~Wu, Q.~Bao, T.~Enoki, Y.~J. Chabal, K.~Ping~Loh, \emph{Nat. Commun.} \textbf{2012}, \emph{3}, 1298.

\bibitem{joshi_precise_2014}
R.~K. Joshi, P.~Carbone, F.~C. Wang, V.~G. Kravets, Y.~Su, I.~V. Grigorieva, H.~A. Wu, A.~K. Geim, R.~R. Nair, \emph{Science} \textbf{2014}, \emph{343}, 752.

\bibitem{eda_chemically_2010}
G.~Eda, M.~Chhowalla, \emph{Adv. Mater.} \textbf{2010}, \emph{22}, 2392.

\bibitem{wu_graphene_2021}
J.~Wu, L.~Jia, Y.~Zhang, Y.~Qu, B.~Jia, D.~J. Moss, \emph{Adv. Mater.} \textbf{2021}, \emph{33}, 2006415.

\bibitem{erickson_determination_2010}
K.~Erickson, R.~Erni, Z.~Lee, N.~Alem, W.~Gannett, A.~Zettl, \emph{Adv. Mater.} \textbf{2010}, \emph{22}, 4467.

\bibitem{dave_chemistry_2016}
S.~H. Dave, C.~Gong, A.~W. Robertson, J.~H. Warner, J.~C. Grossman, \emph{ACS Nano} \textbf{2016}, \emph{10}, 7515.

\bibitem{kumar_impact_2013}
P.~V. Kumar, M.~Bernardi, J.~C. Grossman, \emph{ACS Nano} \textbf{2013}, \emph{7}, 1638.

\bibitem{kumar_scalable_2014}
P.~V. Kumar, N.~M. Bardhan, S.~Tongay, J.~Wu, A.~M. Belcher, J.~C. Grossman, \emph{Nat. Chem.} \textbf{2014}, \emph{6}, 151.

\bibitem{lin_atomistic_2015}
L.-C. Lin, J.~C. Grossman, \emph{Nat. Commun.} \textbf{2015}, \emph{6}, 8335.

\bibitem{williams_computational_2018}
C.~D. Williams, P.~Carbone, F.~R. Siperstein, \emph{Nanoscale} \textbf{2018}, \emph{10}, 1946.

\bibitem{williams_silico_2019}
C.~D. Williams, P.~Carbone, F.~R. Siperstein, \emph{ACS Nano} \textbf{2019}, \emph{13}, 2995.

\bibitem{Futamura2024}
R.~Futamura, T.~Iiyama, T.~Ueda, P.~A. Bonnaud, F.-X. Coudert, A.~Furuse, H.~Tanaka, R.~J.~M. Pellenq, K.~Kaneko, \emph{Nat. Commun.} \textbf{2024}, \emph{15}, 3585.

\bibitem{mouhat_structure_2020}
F.~Mouhat, F.-X. Coudert, M.-L. Bocquet, \emph{Nat. Commun.} \textbf{2020}, \emph{11}, 1566.

\bibitem{Behler2017}
J.~Behler, \emph{Angew. Chem. Int. Ed.} \textbf{2017}, \emph{56}, 12828.

\bibitem{deringer_mlip_2019}
V.~L. Deringer, M.~A. Caro, G.~Csányi, \emph{Adv. Mater.} \textbf{2019}, \emph{31}, 1902765.

\bibitem{Friederich_ff_2021}
P.~Friederich, F.~Häse, J.~Proppe, A.~Aspuru-Guzik, \emph{Nat. Mater.} \textbf{2021}, \emph{20}, 750.

\bibitem{unke_ff_2021}
O.~T. Unke, S.~Chmiela, H.~E. Sauceda, M.~Gastegger, I.~Poltavsky, K.~T. Schütt, A.~Tkatchenko, K.-R. Müller, \emph{Chem. Rev.} \textbf{2021}, \emph{121}, 10142.

\bibitem{thiemann_defect_2021}
F.~L. Thiemann, P.~Rowe, A.~Zen, E.~A. Müller, A.~Michaelides, \emph{Nano Lett.} \textbf{2021}, \emph{21}, 8143.

\bibitem{el-machachi_exploring_2022}
Z.~El-Machachi, M.~Wilson, V.~L. Deringer, \emph{Chem. Sci.} \textbf{2022}, \emph{13}, 13720.

\bibitem{caro_growth_2018}
M.~A. Caro, V.~L. Deringer, J.~Koskinen, T.~Laurila, G.~Cs\'anyi, \emph{Phys. Rev. Lett.} \textbf{2018}, \emph{120}, 166101.

\bibitem{deringer_towards_2018}
V.~L. Deringer, C.~Merlet, Y.~Hu, T.~H. Lee, J.~A. Kattirtzi, O.~Pecher, G.~Csányi, S.~R. Elliott, C.~P. Grey, \emph{Chem. Commun.} \textbf{2018}, \emph{54}, 5988.

\bibitem{wang_structure_2022}
Y.~Wang, Z.~Fan, P.~Qian, T.~Ala-Nissila, M.~A. Caro, \emph{Chem. Mater.} \textbf{2022}, \emph{34}, 617.

\bibitem{ugwumadu_self_2023}
C.~Ugwumadu, R.~Thapa, K.~Nepal, A.~Gautam, Y.~Al-Majali, J.~Trembly, D.~A. Drabold, \emph{J. Chem. Theory Comput.} \textbf{2024}, \emph{20}, 1753.

\bibitem{Li2015}
Z.~Li, J.~R. Kermode, A.~De~Vita, \emph{Phys. Rev. Lett.} \textbf{2015}, \emph{114}, 096405.

\bibitem{Jinnouchi2020}
R.~Jinnouchi, K.~Miwa, F.~Karsai, G.~Kresse, R.~Asahi, \emph{J. Phys. Chem. Lett.} \textbf{2020}, \emph{11}, 6946.

\bibitem{stenczel_castep_gap_2023}
T.~K. Stenczel, Z.~El-Machachi, G.~Liepuoniute, J.~D. Morrow, A.~P. Bartók, M.~I.~J. Probert, G.~Csányi, V.~L. Deringer, \emph{J. Chem. Phys.} \textbf{2023}, \emph{159}, 044803.

\bibitem{clark_castep_2005}
S.~J. Clark, M.~D. Segall, C.~J. Pickard, P.~J. Hasnip, M.~I.~J. Probert, K.~Refson, M.~C. Payne, \emph{Z. Krist.} \textbf{2005}, \emph{220}, 567.

\bibitem{bartok_gap_2010}
A.~P. Bart\'ok, M.~C. Payne, R.~Kondor, G.~Cs\'anyi, \emph{Phys. Rev. Lett.} \textbf{2010}, \emph{104}, 136403.

\bibitem{bartok_soap_2013}
A.~P. Bart\'ok, R.~Kondor, G.~Cs\'anyi, \emph{Phys. Rev. B} \textbf{2013}, \emph{87}, 184115.

\bibitem{deringer_gap_review_2021}
V.~L. Deringer, A.~P. Bartók, N.~Bernstein, D.~M. Wilkins, M.~Ceriotti, G.~Csányi, \emph{Chem. Rev.} \textbf{2021}, \emph{121}, 10073.

\bibitem{batatia_mace_2022}
I.~Batatia, D.~P. Kovacs, G.~Simm, C.~Ortner, G.~Csanyi, {MACE}: Higher Order Equivariant Message Passing Neural Networks for Fast and Accurate Force Fields, in S.~Koyejo, S.~Mohamed, A.~Agarwal, D.~Belgrave, K.~Cho, A.~Oh (Editors), \emph{Advances in Neural Information Processing Systems}, volume~35, Curran Associates, Inc. \textbf{2022} pages 11423--11436.

\bibitem{batatia_design_2022}
I.~Batatia, S.~Batzner, D.~P. Kov\'a{}cs, A.~Musaelian, G.~N.~C. Simm, R.~Drautz, C.~Ortner, B.~Kozinsky, G.~Cs\'a{}nyi, The Design Space of {E(3)}-Equivariant Atom-Centered Interatomic Potentials, arXiv:2205.06643 [stat.ML].

\bibitem{kovacs_evaluation_2023}
D.~P. Kovács, I.~Batatia, E.~S. Arany, G.~Csányi, \emph{J. Chem. Phys.} \textbf{2023}, \emph{159}, 044118.

\bibitem{motevalli_representative_2019}
B.~Motevalli, A.~J. Parker, B.~Sun, A.~S. Barnard, \emph{Nano Futures} \textbf{2019}, \emph{3}, 045001.

\bibitem{McInnes2018}
L.~McInnes, J.~Healy, N.~Saul, L.~Grossberger, \emph{J. Open Source Softw.} \textbf{2018}, \emph{3}, 861.

\bibitem{Cheng2020}
B.~Cheng, R.-R. Griffiths, S.~Wengert, C.~Kunkel, T.~Stenczel, B.~Zhu, V.~L. Deringer, N.~Bernstein, J.~T. Margraf, K.~Reuter, G.~Csanyi, \emph{Acc. Chem. Res.} \textbf{2020}, \emph{53}, 1981.

\bibitem{mattevi_evolution_2009}
C.~Mattevi, G.~Eda, S.~Agnoli, S.~Miller, K.~A. Mkhoyan, O.~Celik, D.~Mastrogiovanni, G.~Granozzi, E.~Garfunkel, M.~Chhowalla, \emph{Adv. Funct. Mater.} \textbf{2009}, \emph{19}, 2577.

\bibitem{tyler_thermal_1953}
W.~W. Tyler, A.~C. Wilson, \emph{Phys. Rev.} \textbf{1953}, \emph{89}, 870.

\bibitem{larsen_ptm_2016}
P.~M. Larsen, S.~Schmidt, J.~Schiøtz, \emph{Model. Simul. Mater. Sci. Eng.} \textbf{2016}, \emph{24}, 055007.

\bibitem{OVITO}
A.~Stukowski, \emph{Model. Simul. Mater. Sci. Eng.} \textbf{2009}, \emph{18}, 015012.

\bibitem{maphiri_novel_2021}
V.~M. Maphiri, G.~Rutavi, N.~F. Sylla, S.~A. Adewinbi, O.~Fasakin, N.~Manyala, \emph{Nanomaterials} \textbf{2021}, \emph{11}, 1909.

\bibitem{DeTomas2017}
C.~de~Tomas, I.~Suarez-Martinez, F.~Vallejos-Burgos, M.~J. L\'o{}pez, K.~Kaneko, N.~A. Marks, \emph{Carbon} \textbf{2017}, \emph{119}, 1.

\bibitem{Reyes2022}
M.~Ceniceros-Reyes, K.~Marín-Hernández, U.~Sierra, E.~Saucedo-Salazar, R.~Mendoza-Resendez, C.~Luna, P.~Hernández-Belmares, O.~Rodríguez-Fernández, S.~Fernández-Tavizón, E.~Hernández-Hernández, E.~D. Barriga-Castro, \emph{Surf. Interfaces} \textbf{2022}, \emph{35}, 102448.

\bibitem{valentini_tuning_2023}
C.~Valentini, V.~Montes-García, P.~A. Livio, T.~Chudziak, J.~Raya, A.~Ciesielski, P.~Samorì, \emph{Nanoscale} \textbf{2023}, \emph{15}, 5743.

\bibitem{Nguyen_rgo_2015}
N.~D.~K. Tu, J.~Choi, C.~R. Park, H.~Kim, \emph{Chem. Mater.} \textbf{2015}, \emph{27}, 7362.

\bibitem{Pelaez_2021}
M.~Pelaez-Fernandez, A.~Bermejo, A.~Benito, W.~Maser, R.~Arenal, \emph{Carbon} \textbf{2021}, \emph{178}, 477.

\bibitem{golze_accurate_2022}
D.~Golze, M.~Hirvensalo, P.~Hernández-León, A.~Aarva, J.~Etula, T.~Susi, P.~Rinke, T.~Laurila, M.~A. Caro, \emph{Chem. Mater.} \textbf{2022}, \emph{34}, 6240.

\bibitem{Zarrouk2024experimentdriven}
T.~Zarrouk, R.~Ibragimova, A.~P. Bartók, M.~A. Caro, \emph{J. Am. Chem. Soc.} \textbf{2024}, \emph{Article ASAP}, DOI: 10.1021/jacs.4c01897.

\bibitem{aarva_xray_2021}
A.~Aarva, S.~Sainio, V.~L. Deringer, M.~A. Caro, T.~Laurila, \emph{J. Phys. Chem. C} \textbf{2021}, \emph{125}, 18234.

\bibitem{Kovacs2023a}
D.~P. Kov\'a{}cs, J.~H. Moore, N.~J. Browning, I.~Batatia, J.~T. Horton, V.~Kapil, W.~C. Witt, I.-B. Magd\u{a}u, D.~J. Cole, G.~Cs\'a{}nyi, {MACE}-{OFF23}: {Transferable} {Machine} {Learning} {Force} {Fields} for {Organic} {Molecules}, arXiv:2312.15211 [physics.chem-ph].

\bibitem{Batatia2023}
I.~Batatia, P.~Benner, Y.~Chiang, A.~M. Elena, D.~P. Kovács, J.~Riebesell, X.~R. Advincula, M.~Asta, M.~Avaylon, W.~J. Baldwin, F.~Berger, N.~Bernstein, A.~Bhowmik, S.~M. Blau, V.~Cărare, J.~P. Darby, S.~De, F.~Della~Pia, V.~L. Deringer, R.~Elijošius, Z.~El-Machachi, F.~Falcioni, E.~Fako, A.~C. Ferrari, A.~Genreith-Schriever, J.~George, R.~E.~A. Goodall, C.~P. Grey, P.~Grigorev, S.~Han, W.~Handley, H.~H. Heenen, K.~Hermansson, C.~Holm, J.~Jaafar, S.~Hofmann, K.~S. Jakob, H.~Jung, V.~Kapil, A.~D. Kaplan, N.~Karimitari, J.~R. Kermode, N.~Kroupa, J.~Kullgren, M.~C. Kuner, D.~Kuryla, G.~Liepuoniute, J.~T. Margraf, I.-B. Magdău, A.~Michaelides, J.~H. Moore, A.~A. Naik, S.~P. Niblett, S.~W. Norwood, N.~O'Neill, C.~Ortner, K.~A. Persson, K.~Reuter, A.~S. Rosen, L.~L. Schaaf, C.~Schran, B.~X. Shi, E.~Sivonxay, T.~K. Stenczel, V.~Svahn, C.~Sutton, T.~D. Swinburne, J.~Tilly, C.~van~der Oord, E.~Varga-Umbrich, T.~Vegge, M.~Vondrák, Y.~Wang, W.~C. Witt, F.~Zills, G.~Csányi, A foundation model for atomistic
  materials chemistry, arXiv:2401.00096 [physics.chem-ph].

\bibitem{zheng_swelling_2017}
S.~Zheng, Q.~Tu, J.~J. Urban, S.~Li, B.~Mi, \emph{ACS Nano} \textbf{2017}, \emph{11}, 6440.

\bibitem{kapil_first_2022}
V.~Kapil, C.~Schran, A.~Zen, J.~Chen, C.~J. Pickard, A.~Michaelides, \emph{Nature} \textbf{2022}, \emph{609}, 512.

\end{thebibliography}
\end{document}


\title{\Large {\bf Supporting Information for}\\[4mm] ``Accelerated First-Principles Exploration of Structure and Reactivity in Graphene Oxide''}

\author[1]{Zakariya El-Machachi}
\author[1]{Damyan Frantzov}
\author[1]{A. Nijamudheen}
\author[2]{Tigany Zarrouk}
\author[2]{Miguel A. Caro}
\author[1]{Volker L. Deringer\thanks{volker.deringer@chem.ox.ac.uk}}

\affil[1]{Inorganic Chemistry Laboratory, Department of Chemistry, University of Oxford, Oxford OX1 3QR, United Kingdom}
\affil[2]{Department of Chemistry and Materials Science, Aalto University, 02150 Espoo, Finland}

\date{}

\maketitle

\thispagestyle{empty}

\clearpage
\setstretch{1.5}

\section*{Computational methods}

{\bf ML acceleration for CASTEP.} 
The CASTEP+ML scheme described in Ref.\ \citenum{stenczel_castep_gap_2023} was used to accelerate the sampling of the relevant configurational space (as compared to full {\em ab initio} molecular dynamics) and thus to construct an initial dataset for ML potential fitting. All DFT computations were performed at the $\Gamma$ point and used the PBE functional with a plane-wave cutoff of 550 eV and a SCF halting criterion of $\Delta E < 10^{-5}$ eV at.$^{-1}$. A Gaussian smearing width of 0.1 eV was applied and all computations performed are non-spin-polarized. 

For CASTEP+ML, the adaptive fitting method (increasing and decreasing the number of steps between DFT checks, $n$, depending on model performance; Ref.\ \citenum{stenczel_castep_gap_2023}) was used, where $n_{\text{min}}=1$ and $n_{\text{max}}=10000$ with an adaptive scaling factor of 2. The tolerances used for refitting checks were an energy difference of $< 0.01$ eV at.$^{-1}$, a maximum force difference $< 3.00$ eV Å$^{-1}$, and a force RMSE of $<0.50$ eV Å$^{-1}$. 

The GAP models fitted by CASTEP+ML are based on a combination of 2-body, 3-body, and SOAP terms, similar to the C-GAP-17 model \cite{Deringer2017GAP}, however with modified hyperparameters and much fewer sparse (representative) points particularly for the SOAP term. The GAP fitting string used for on-the-fly potential fitting is given in Listing 1 below.

\begin{lstlisting}[caption=GAP fitting string for CASTEP+ML runs.]
  default_sigma: "0.008 0.04 0 0"
  descriptor_str: "distance_Nb order=2 cutoff=4.5 covariance_type=ard_se delta=2.0 theta_uniform=1.0 sparse_method=uniform add_species=T n_sparse=15 : distance_Nb order=3 cutoff=2.8 covariance_type=ard_se delta=0.5 theta_uniform=1.0 add_species=T n_sparse=50 sparse_method=uniform : soap cutoff=4.5 covariance_type=dot_product zeta=4.0 delta=0.05 atom_sigma=0.5 l_max=8 n_max=8 n_sparse=200 sparse_method=cur_points"
  extra_gap_opts: "sparse_jitter = 1.0e-8"
\end{lstlisting}

{\bf MACE fitting.}
Equivariant MACE models \cite{batatia_mace_2022} were fitted on a single NVIDIA RTX A6000 graphics card in a Linux workstation. The final model required about 53 hours to train. 
We used the MACE code version 0.2.0, development branch, commit {\tt 55f7411}.
The input used for MACE fitting is given in Listing 2 below.

\clearpage 

\begin{lstlisting}[caption=MACE fitting input. ``x'' denotes the iteration number in the training and testing file.]
    CUDA_VISIBLE_DEVICES="$gpu_id" python mace/scripts/run_train.py \
    --name="MACE_model" \
    --train_file="structures/iter-x-train.xyz" \
    --valid_fraction=0.10 \
    --test_file="structures/iter-x-test.xyz" \
    --config_type_weights='{"Default":1.0}' \
    --model="ScaleShiftMACE" \
    --hidden_irreps='128x0e+128x1o' \
    --loss='huber' \
    --r_max=3.7 \
    --batch_size=30 \
    --max_num_epochs=2000 \
    --swa \
    --default_dtype='float32' \
    --energy_key='QM_energy' \
    --forces_key='QM_forces' \
    --stress_key=None \
    --start_swa=1000 \
    --swa_energy_weight=1000 \
    --swa_forces_weight=100 \
    --lr=0.001 \
    --ema \
    --ema_decay=0.99 \
    --amsgrad \
    --restart_latest \
    --device=cuda \
    --seed=123 \
\end{lstlisting}

\clearpage 

\begin{figure}[t]
    \centering
    \includegraphics[width=0.9\textwidth]{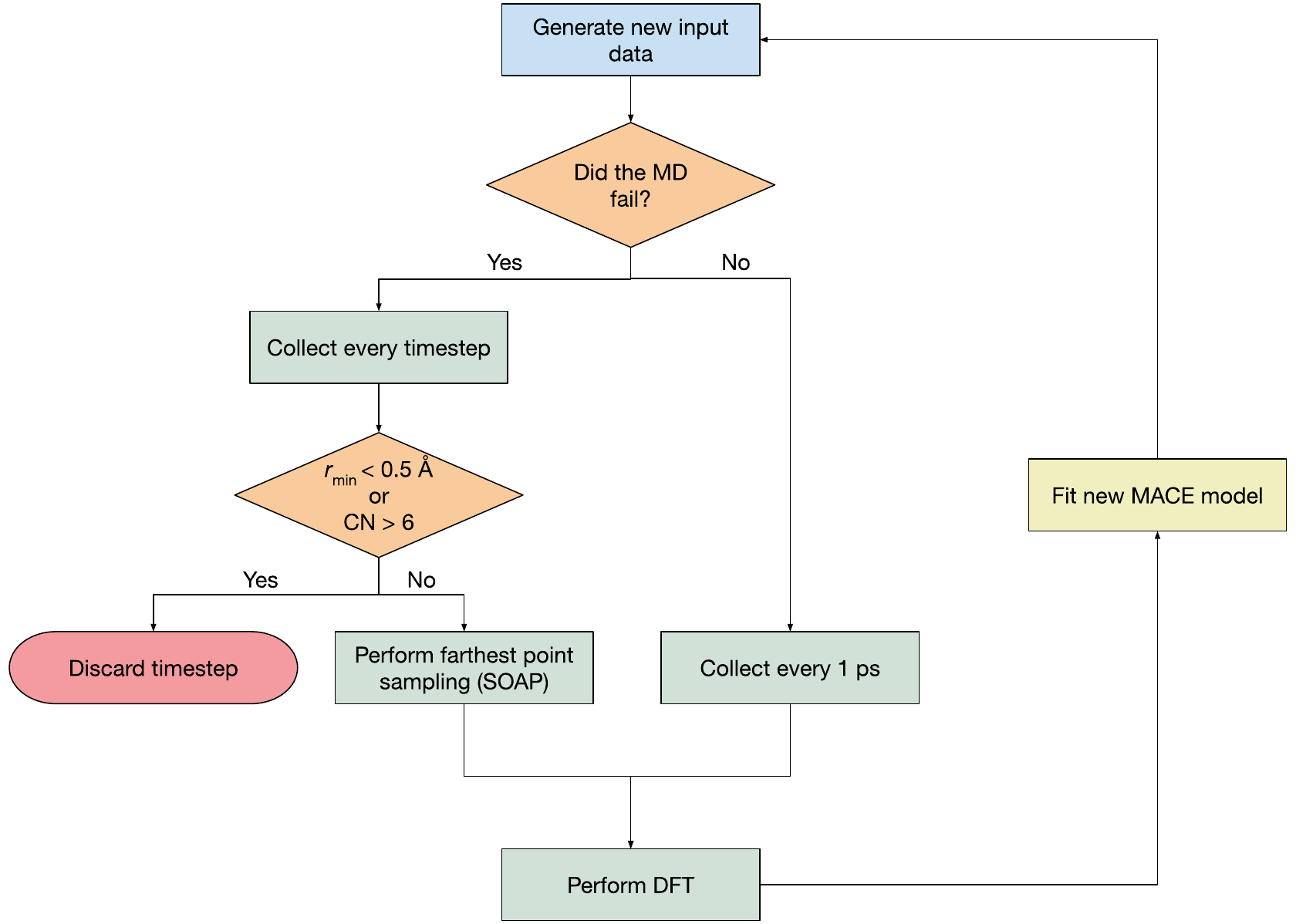}
    \caption{Flowchart outlining the process for selecting structures during iterative training.}
    \label{fig:filtering}
\end{figure}

{\bf Data filtering.} The iterative process occasionally generated structures with very (even unreasonably) high energies and forces. We found that a process for filtering high-energy and -force structures throughout the iterative training process was important for the stability of early MACE models. Including highly unfavourable structures resulted in a less smooth regression---however, without inclusion of these unfavourable structures, some early models displayed instabilities and led to catastrophic failure during molecular dynamics (MD), such as lost atoms or nonphysical clustering. The approach to filtering structures during iterative training is illustrated in Figure \ref{fig:filtering}.

Initially, we check whether a given MD run has completed without failing, in which case we simply collect structures at every 1 ps from the MD trajectory. However, if a run did fail, we investigate the trajectory in more detail. We heuristically define two criteria to determine whether a structure is fully nonphysical from the iterative MD (and should therefore not be included in the training dataset). We deem a structure to be fully nonphysical if it meets one of the following criteria:
\begin{itemize}
    \item Any two atoms come closer than 0.5 \AA{} (denoted ``$r_\text{min}$'' in Figure \ref{fig:filtering})
    \item The coordination number (CN) of any atom is greater than 6
\end{itemize}
The cutoff values were experimented with and proved to be sensitive to the types of structures filtered. Values of $> 0.5$ Å did not include enough unfavourable structures, whereas values of $< 0.5$ Å would include structures that then failed to converge in DFT. 

Farthest point sampling (FPS) was used on the structures from the trajectory that remained after the filtering step. FPS was performed on the per-structure SOAP vectors constructed from the average atomic SOAP vectors of that structure. This enabled an automated and rigorous sampling of configurational space related to these failed MD runs. The final dataset was filtered of structures containing force components above 50 eV/Å.

{\bf Iterative training.}
MD simulations using the Atomic Simulation Environment (ASE) were used to generate configurations iteratively. The Nosé–Hoover thermostat was used with an MD timestep of 0.2 fs. For the first four iterations, the temperature of the MD simulation was increased in increments of 300 K, starting from 600 K where model-1, which was trained on the CASTEP+ML data generated at 300 K, was used to drive MD across the parameter space and ending at 1,500 K for model-4, which had been trained iteratively from data generated from subsequent models. This iterative approach was chosen to enable a wider-ranging sampling of configuration space, without jeopardising stability in early models. We found that up to 1,200 K, the resulting MACE potentials were stable for at least 10 ps. At 1,200 K, the MACE potential at this iteration (model-3) was not stable for two runs at high O content and low OH/O ratio ($p_{1}=0.40\text{, } p_{2} =0.00 \text{ and } p_{1}=0.50\text{, }  p_{2} =0.25$), where model-3 incorrectly predicts a minimum in the potential energy surface (PES), leading to unphysical clustering of atoms. This behaviour is somewhat expected during early stages of the iterative training of ML potentials, where the configurational space is still actively being explored. Furthermore, explicit models of isolated dimers are not included in the training data and thus extremely short-range interactions ($r \le 0.5 $ Å) are poorly described by early models. This was also observed at 1,500 K in the first instance (iteration 4) where two configurations failed ($p_{1}=0.50, p_{2} =0.00 \text{ and } p_{1}=0.50, p_{2} =0.25$). 7 structures were removed from the final training dataset and 0 were removed from the test set.
 
\begin{figure}
    \centering
    \includegraphics[width=0.9\textwidth]{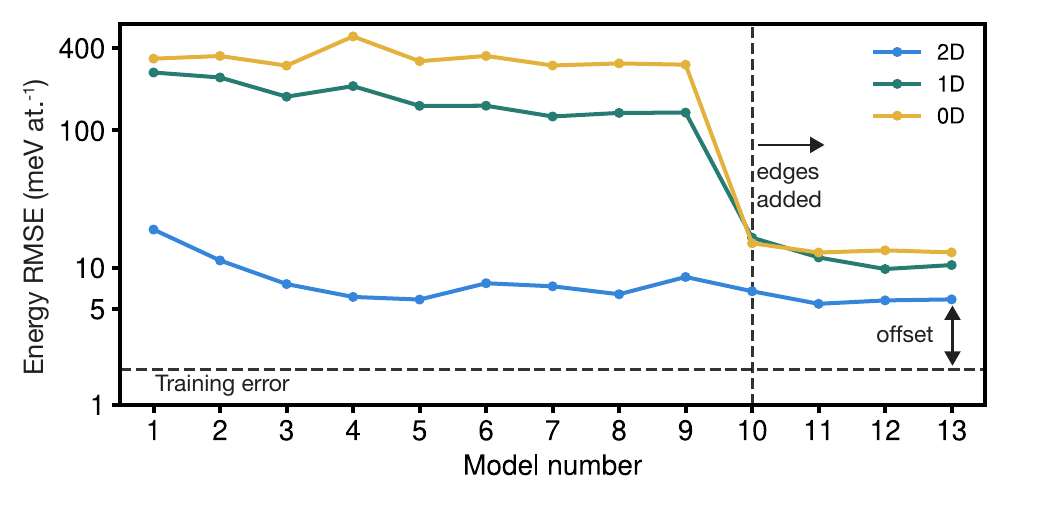}
    \caption{Energy RMSE for the external test sets discussed in the main text. The training error is shown as a dashed line at 1.8 meV at.$^{-1}$. }
    \label{fig:energy-rmse}
\end{figure}

\begin{figure}
    \centering
    \includegraphics[width=0.9\textwidth]{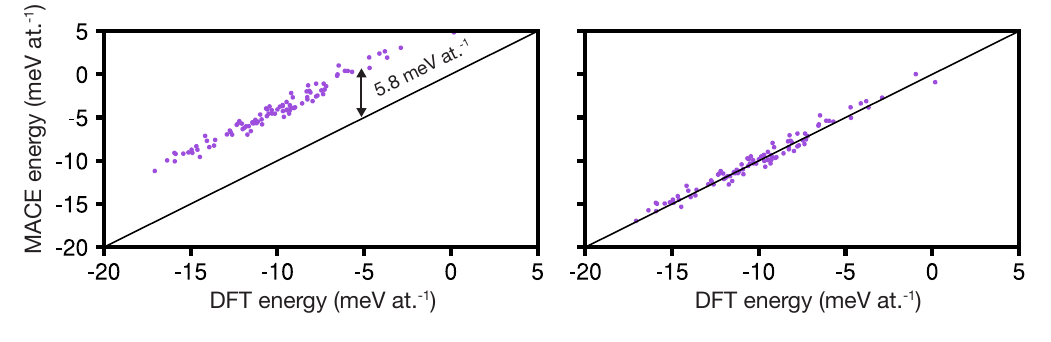}
    \caption{Energy parity plot for the 2D external test set. The left plot is the raw data, the right plot subtracts the shift to centre the plot.}
    \label{fig:scatter-energy}
\end{figure}

\begin{figure}
    \centering
    \includegraphics[width=0.9\textwidth]{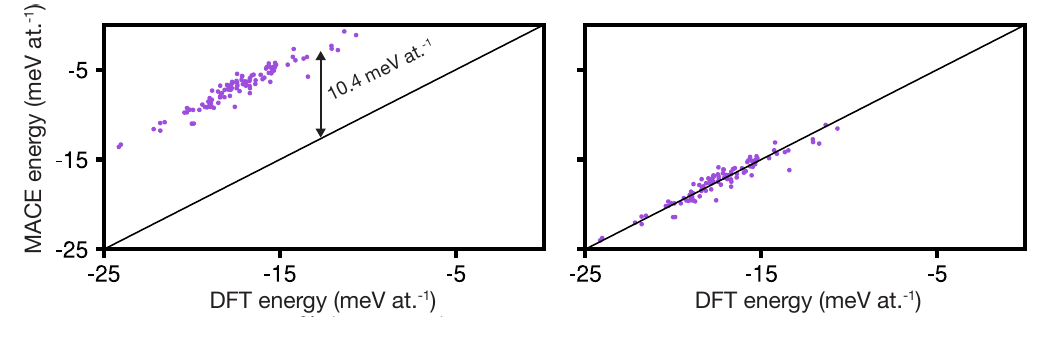}
    \caption{Energy parity plot for the 1D external test set. The left plot is the raw data, the right plot subtracts the shift to centre the plot.}
    \label{fig:scatter-1D}
\end{figure}

\begin{figure}
    \centering
    \includegraphics[width=0.9\textwidth]{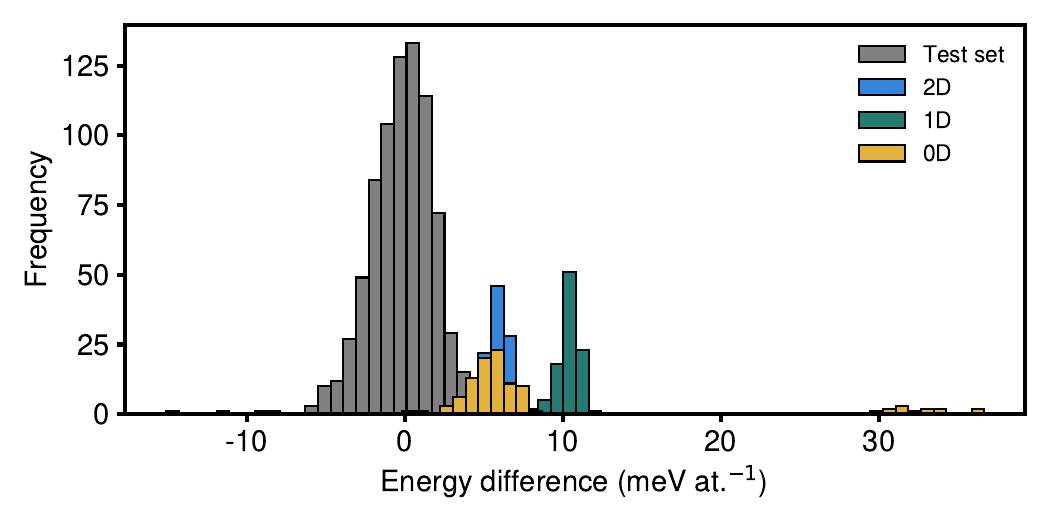}
    \caption{Histograms showing energy difference between final MACE model and DFT on the internal test set (grey); external AIMD 2D (blue) and 1D (green) test set and external 0D (yellow) test set using structures from Ref.~\citenum{motevalli_representative_2019}.}
    \label{fig:hist-train}
\end{figure}

{\bf Energy RMSE.}
The energy RMSEs of the MACE models are characterised in Figures \ref{fig:energy-rmse} to \ref{fig:hist-train}. Briefly, the energy predictions are offset by a notable amount, shifting the energy predictions up by approximately 6 meV at.$^{-1}$ for the external AIMD 2D test set and approximately 10 meV at.$^{-1}$ for the external AIMD 1D test set. The external 0D case (structures taken from Ref.~\citenum{motevalli_representative_2019}) is not shown in the parity plots, as the structures have varying numbers of atoms, however the highest error is approximately 30 meV at.$^{-1}$ with the average being approximately 8 meV/atom as seen in Figure \ref{fig:hist-train}. 

We believe that these offsets do not meaningfully impact the potentials' performance in dynamics since only the {\em relative} positions of minima and maxima are significant, not their absolute values. The variance of the data is perhaps another useful metric here, however this is not typically reported and thus we will refrain from doing so. However, it is important to note when reporting the energy RMSE, these potentials appear to perform worse than they actually do. We do not think that this offset is an issue specific to the MACE framework, and we will continue to investigate its origin.

At the end of the process, our final MACE model yielded an accuracy of 0.174 kJ mol$^{-1}$ (1.8 meV at.$^{-1}$) for energies on the train set and 98.1 meV Å$^{-1}$ for forces on the train set, the latter being in close agreement with the results for the external test set. 

We also report the energy difference for the ``internal'' test set in Figure \ref{fig:hist-train}, which are defined as structures generated by the MACE model during iterative MD but not included in the training. In this internal test set, we have 800 structures and observe that there is no offset present as the data are centered around 0 meV at.$^{-1}$. As mentioned earlier, the 0D data show two regions, one centred at $\approx$ 8 meV at.$^{-1}$, which corresponds to H-terminated edges, and another region centred at $\approx$ 33 meV at.$^{-1}$, which corresponds to edges with no H atoms. 

{\bf Polyhedral template matching.}
Polyhedral Template Matching (PTM) is used to identify atoms resembling simple crystalline structures at a local level \cite{larsen_ptm_2016}. Within PTM, each input particle establishes a correlation between its local environment and the template of choice. If a correlation is identified, an RMSD (Root Mean Square Deviation) value is computed to measure spatial deviation from the ideal structural template, determining the match quality. In this work we used an RMSD cutoff value of 0.15 to determine whether a given local environment was ``graphenic'' or not. This method was used as implemented in OVITO 3.6.0 \cite{OVITO}.

{\bf Ring statistics.}
Ring statistics were determined using a shortest-path algorithm \cite{Franzblau1991ring} as implemented in \texttt{matscipy} \cite{matscipy, Grigorev2024}.

{\bf Cell optimisation.}
The structures from the production runs were cooled down to 300 K over 100 ps and then underwent a full cell relaxation. This was done within ASE using the \lstinline{FrechetCellFilter} class and the LBFGS optimizer. The force threshold was 10 meV at.$^{-1}$.

{\bf Methodology for X-ray photoelectron spectroscopy.}
The GW-corrected delta Kohn--Sham core-electron binding energy model from Ref.~\citenum{golze_accurate_2022} was used to predict the C 1s core-electron binding energies (CEBE). This is a single \lstinline{soap_turbo} \cite{Caro2019soap} model which is trained on neutral bulk delta Kohn--Sham CEBEs corrected by the difference of a GW core-electron binding energy and a charged delta Kohn--Sham core-electron binding energy on a configuration ``carved'' from an extended structure: 
\begin{equation}
  \Delta \text{KS}^0_{\rm ext} + GW_{\rm carv} - \text{KS}^{+}_{\rm carv}.  
\end{equation}
Details of the methodology are given in Ref.\ \citenum{golze_accurate_2022}.

Deconvolutions were performed by considering each structure at a particular timestep as an undirected graph, where edges are bonds, and nodes are atoms differentiated by species. Bonds cutoffs were defined in ASE using \lstinline{natural_cutoffs(atoms, mult=1.2)}. A local subgraph was made for each carbon atom by adding edges and nodes initially from the atom's first neighbors, and if oxygen was found, adding the additional edges and nodes from the oxygen neighbors (which is necessary to discern epoxide and ether groups). Motifs were found by seeking subgraph isomorphisms between a reference dataset of motif graphs (e.g. aldehyde, alcohol, ketone etc). These were sought hierarchically, in the order: carboxylic acid, aldehyde, alcohol, carbonate, peroxide, ester, epoxide, ether, ketone, \ce{CH}, \ce{CO2}, \ce{CO}, sp$^{3}$, sp$^{2}$, sp.

\clearpage

\section*{Supplementary results}

\subsection*{Mass-loss profiles (supplement to Figure 3b)}
\begin{figure}[h]
    \centering
    \includegraphics[width=0.9\textwidth]{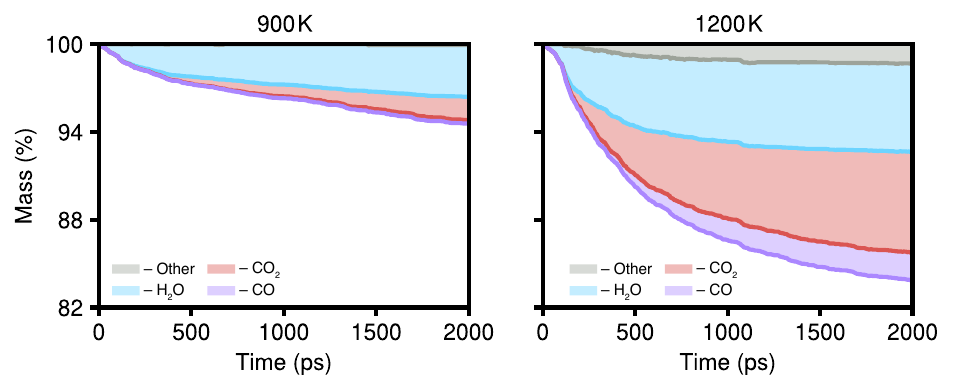}
    \caption{Mass loss profiles for the 900 K and 1,200 K simulations.}
    \label{fig:SI-mass}
\end{figure}

We carried out two further annealing simulations in parallel, starting from the same large-scale structural model as described in the main text, but now annealing at 900 and 1,200 K, respectively. The mass loss profiles for those lower-temperature simulations are shown in Figure \ref{fig:SI-mass}. Overall, the change in mass at 1,200 K begins to plateau, converging at approximately 17\%.  
This mass loss profile agrees qualitatively with experimental thermogravimetric data from Ref.\ \cite{valentini_tuning_2023}, where at the start of the thermal reduction, the mass loss follows a rapid exponential curve before reaching a linear phase for the remainder of the reduction process. In this way it is qualitatively similar to the 1,500 K simulation reported in the main text. The 900 K annealing run (left-hand side of Figure \ref{fig:SI-mass}), however, shows much a much less pronounced mass loss which is largely due to loss of \ce{H2O}.

\subsection*{Graphene-like atoms and 6-membered rings (supplement to Figure 3c)}
\begin{figure}
    \centering
    \includegraphics[width=0.9\textwidth]{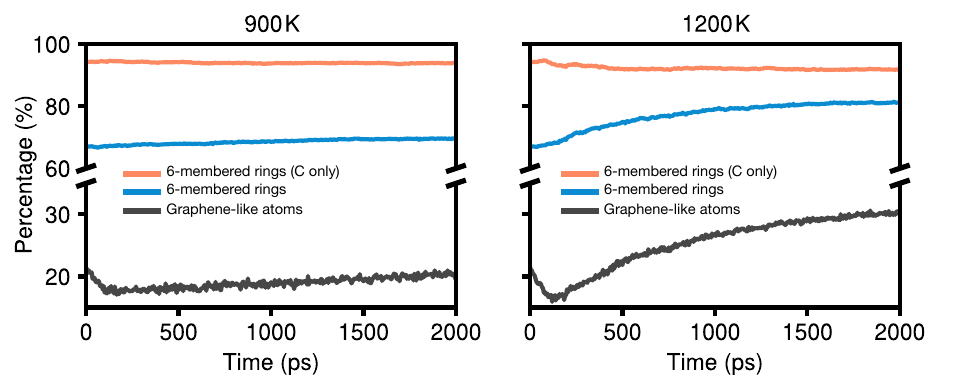}
    \caption{Graphene-like atoms and 6-membered rings profiles for the 900 K and 1,200 K simulations.}
    \label{fig:SI-rings}
\end{figure}

Figure \ref{fig:SI-rings} shows the evolution of the count of graphene-like motifs and 6-membered rings in the 900 and 1,200 K simulations. In our simulations, the ``graphene-like'' content eventually  reaches $\approx 40$ \% for the 1,500 K MD run (Figure 3c) whereas it is still increasing for the 1,200 K run at just over 30 \% (Figure \ref{fig:SI-rings}). 900 K appears to be too low of a temperature to effectively reduce GO in nanosecond simulations: the degree of ``graphene-like'' similarity remains essentially steady, with only a very slight increase over 2 ns.
The fraction of ``graphene-like'' atoms (dark grey) in combination with the percentage of 6-membered rings (blue) provides insight into the degree of crystallinity attained during the annealing process. We also show the percentage of 6-membered rings with only carbon atoms (orange) which is naturally quite high, but does not indicate how graphitic the structure is. 

Annealing temperatures of 900 K appear to have no significant effect on the percentage of 6-membered rings which stays approximately constant. In addition to this, the count of ``graphene-like'' atoms decreases slightly during the initial heating phase. This is then followed by a very gentle linear increase which is still less than the initial percentage.

At 1,200 K, we observe more dynamic behaviour, where the percentage of 6-membered rings increases whereas that of 6-membered rings containing only carbon decreases, suggesting that more oxygen atoms are incorporated into the graphene network. This is supported by the X-ray photoelectron spectroscopy (XPS) data seen in Figure \ref{fig:SI-XPS}. In addition to this, the percentage of ``graphene-like'' atoms decreases at a faster rate when compared to the 900 K data over the heating period, before increasing at a faster rate before beginning to plateau after 2 ns. Overall, we see the percentage of ``graphene-like'' atoms increase with that of the 6-membered rings, as expected.

\subsection*{Functional groups (supplement to Figure 3d)}
Tracking the functional groups during the annealing process provides insight into the mechanisms by which the structure becomes more graphene-like. For the 900 K simulation, there is a gradual decrease of $\text{sp}^{3}$ groups (dark and light blue) and a corresponding gradual increase of $\text{sp}^{2}$ groups (orange, red, purple). Overall, this profile further emphasises that 900 K is too low a temperature to effectively anneal GO structural models on the ns timescale using our model.

\begin{figure}
    \centering
    \includegraphics[width=0.9\textwidth]{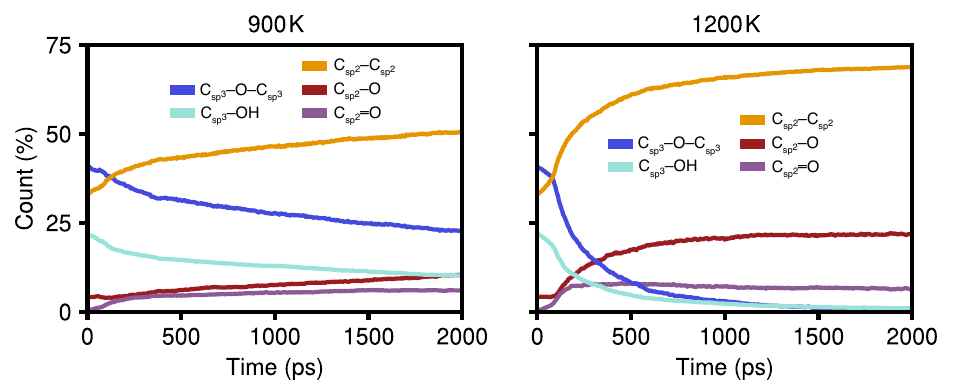}
    \caption{Functional group profiles for the 900 K and 1,200 K simulations.}
    \label{fig:SI-func}
\end{figure}

In contrast, the 1,200 K simulation follows a similar profile to the 1,500 K simulation, albeit at a slower rate. Here it is clear that $\text{sp}^{3}$ groups are nearly completely removed after 2 ns with a similar growth profile of $\text{sp}^{2}$ groups, including a characteristic steady-state period for C$_{\text{sp}^{2}}$–O groups. 

\begin{figure}
    \centering
    \includegraphics[width=0.9\textwidth]{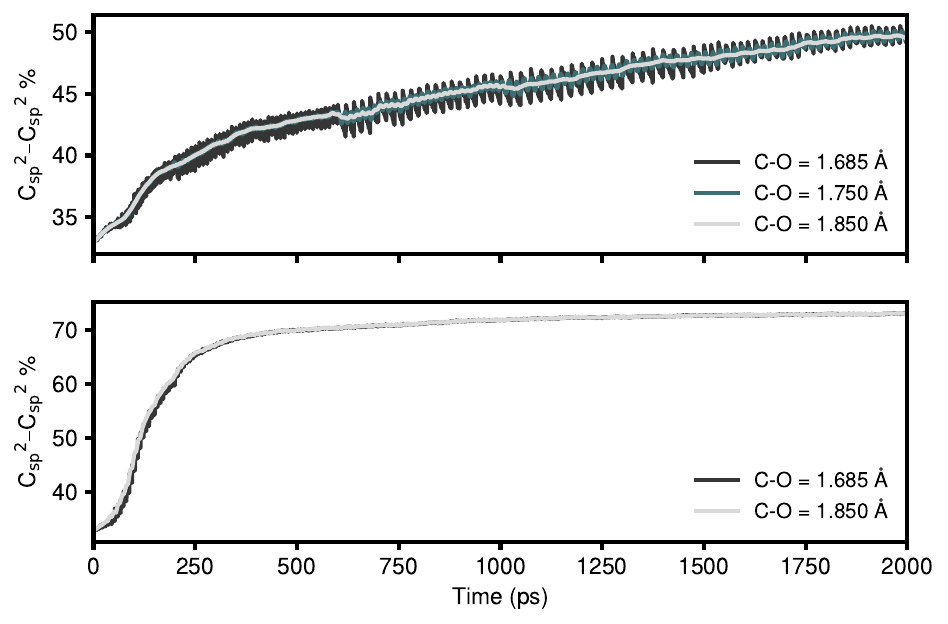}
    \caption{{\em Top:} Oscillations for the C$_{\text{sp}^{2}}$-C$_{\text{sp}^{2}}$ count at 900 K, as evaluated using different C–O bond lengths for the cutoff definition. {\em Bottom:} Oscillations at 1,500 K. The data with a cutoff value of 1.850 \AA{} are shown in Figure 3d of the main text.}
    \label{fig:SI-oscillation}
\end{figure}

We note that using the average ionic radius for the C–O bond (C$_{\text{ionic}}$/2 + 
O$_{\text{ionic}}$/2) gives oscillatory behaviour for the C$_{\text{sp}^{2}}$-C$_{\text{sp}^{2}}$ count, as indicated by the black line in the top panel of Figure \ref{fig:SI-oscillation}. As the defined bond cutoff is increased beyond the ionic radius (from 1.685 to 1.750 to 1.850 \AA{}), we see the magnitude of oscillations decrease. We observe changes in the oscillatory behaviour during the course of the simulation, which may be due to gaseous species being removed and simulations restarted at some points. Much less pronounced oscillations are seen for the 1,500 K simulation, which is discussed in the main text (Figures \ref{fig:SI-oscillation} and 3d). For other species, all structural analysis was performed using the ionic-radius-based cutoff values. 

\subsection*{X-ray photoelectron spectroscopy (supplement to Figure 4)}
After the structures had been annealed, they were cooled down to 300 K over 100 ps and then underwent a full cell optimisation to relax the structure. XPS calculations were performed on these fully relaxed structures along with selected structures along the trajectory. 

\begin{figure}[hb]
    \centering
    \includegraphics[width=0.9\textwidth]{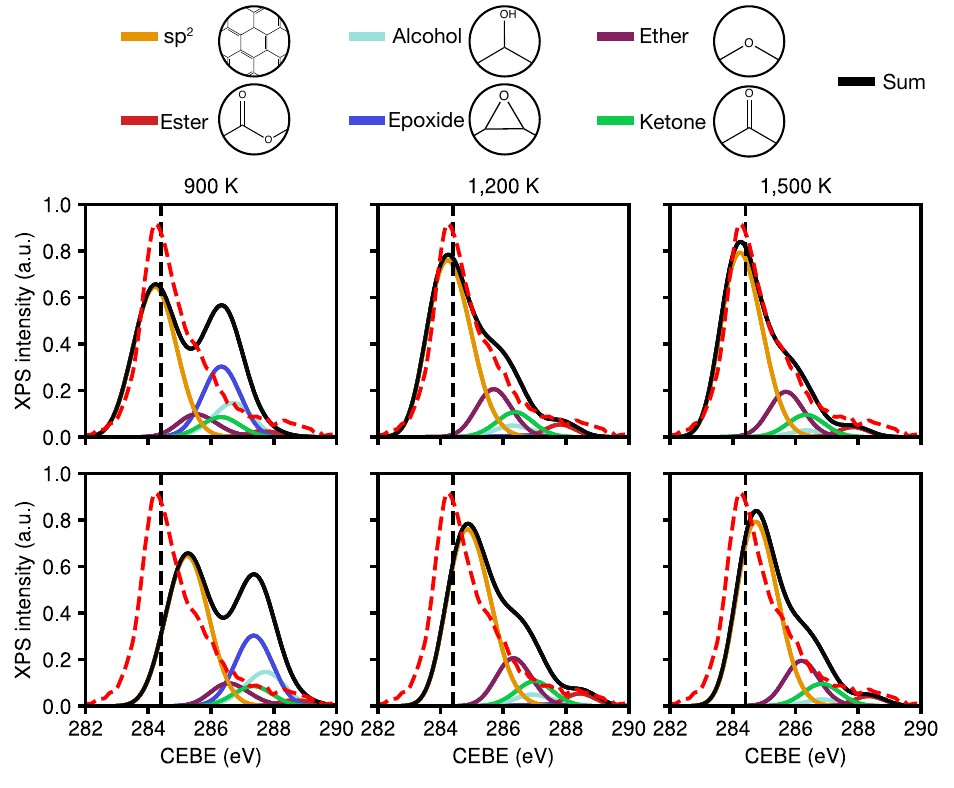}
    \caption{ML-model-predicted XPS spectra for the 900 K, 1,200 K, and 1,500 K simulations. \textbf{Top:} Shifted XPS spectra to experimental sp$^{2}$ reference value at 284.4 eV. The red dashed line shows experimental data taken from Ref.\ \citenum{valentini_tuning_2023}. \textbf{Bottom:} Unshifted XPS data showing the effect of electronegativity on CEBE values.}
    \label{fig:SI-XPS}
\end{figure}

We summarize additional results in Figure \ref{fig:SI-XPS}. The top panel shows the shifted simulated spectra, where the shift has been introduced to better fit the experimental data (red, dashed). This is done since deconvolutions of experimental data are fixed at reference values (sp$^{2}$ = 284.4 eV, black dashed lines) to perform this task. In contrast, the simulated spectra are not constrained by this, and so we report both types of data in Figure \ref{fig:SI-XPS}; the data in the top panel are shifted towards the reference peak and the bottom panel shows the un-shifted simulation data. The experimental data taken from Ref.~\citenum{valentini_tuning_2023} have had their background removed and have been normalised in order to be compared to simulated spectra. 

Both approaches provide useful information---the former giving comparison to experimental data, the latter providing information about the shifting of the peak towards the reference sp$^{2}$ value as oxygen is removed from the system after annealing at different temperatures. We have chosen to display only the six most important groups as other functional groups did not contribute substantially. 

At 900 K, we see a spectrum which appears very different to the experimental data from Ref.~\citenum{valentini_tuning_2023} (red dashed line). This is due to the increased presence of oxygen groups in the structure, in particular, epoxide and alcohol groups which are yet to be removed. There are further contributions from ether, ester, and ketone groups. There is also a shift to higher CEBEs due to the increased electronegativity from the oxygen groups present. 

At 1,200 K, the oxygen peak has now been reduced to a shoulder peak with the epoxide peak being nearly removed entirely and the alcohol peak being greatly lowered. In contrast the ether group, ketone group, and ester group peaks have all increased greatly. This agrees with the wider trend of transitioning from sp$^{3}$ to sp$^{2}$. The shift down to the sp$^{2}$ reference peak is also observed in the bottom panel due to fewer oxygen groups being present and a greater presence of sp$^{2}$ environments. 

Finally, at 1,500 K, we observe the best agreement with the experimental XPS data of Ref.~\citenum{valentini_tuning_2023}. There are no significant contributions from epoxide environments and, in total, fewer oxygen groups present. This can be clearly seen by the reduction in the shoulder peak, giving closer agreement experiment than for the two other simulations at lower temperatures. Additionally, there is a closer shift to the reference sp$^{2}$ value, further highlighting the removal of oxygen. The results in Figure \ref{fig:SI-XPS} therefore provide further justification that 1,500 K is an appropriate choice of annealing temperature for the simulations reported in the main text.

\clearpage 
\section*{Supplementary references}